\documentclass[a4paper, 11pt]{article}

\usepackage[utf8]{inputenc} 
\usepackage[T1]{fontenc}    
\usepackage[english]{babel} 
\usepackage{amsmath}        
\usepackage{graphicx}       
\usepackage[margin=1in]{geometry} 

\usepackage{siunitx}

\usepackage{hyperref}

\usepackage{cleveref}
\hypersetup{
    colorlinks=true,
    linkcolor=blue,
    filecolor=magenta,      
    urlcolor=cyan,
    pdftitle={Your Paper Title},
    pdfpagemode=FullScreen,
}

\usepackage[utf8]{inputenc}
\usepackage[T1]{fontenc}
\usepackage{xcolor} 
\usepackage{authblk}

\usepackage[normalem]{ulem}

\title{\textbf{First Dark Photon Search Results from the Dandelion Experiment}}

\author[1]{I. Ourahou}
\author[1]{S. Savorgnano}
\author[1]{C. Beaufort}
\author[2,1]{M. Bastero-Gil}
\author[1]{J. Bounmy}
\author[1]{A. Catalano}
\author[1]{J. Macias-Perez} 
\author[1]{D. Santos}
\author[1]{C. Smith}
\author[1]{F. Naraghi}
\author[1]{D.~Tourres}
\author[1]{F. Vezzu}

\affil[1]{Laboratoire de Physique Subatomique et Cosmologie, Universit\'{e} Grenoble-Alpes, CNRS/IN2P3, 38000 Grenoble, France}

\affil[2]{Departamento de Física Teórica y del Cosmos, Universidad de Granada,
Granada-18071, Spain}

\affil[ ]{E-mail: \textcolor{violet}{ilias.ourahou@lpsc.in2p3.fr} , \textcolor{violet}{daniel.santos@lpsc.in2p3.fr} }

\date{}

\begin{document}

\maketitle
\vspace{-1 cm}
\begin{abstract}
This paper presents the first results from the Dandelion experiment, a directional detection experiment, which searches for 1 meV dark photon dark matter. We use a spherical mirror to convert dark photons into standard millimeter-wavelength photons that can then be detected with an array of 221 Kinetic Inductance Detectors (KIDs) cooled down to 150 mK within the KISS-NIKA camera (KIDs Interferometric Spectral Surveyor) and operating between 150 and 350 GHz. We used 1480 minutes of data to search for the signal  of dark photons in the KID detectors, which is expected to be modulated due to the Earth's rotation. Our main challenge was to deal with a large background from room temperature and stray-light fluctuations. We used a de-correlation analysis to remove these background fluctuations. Templates of the background fluctuations were constructed from a Principal Component Analysis decomposition of detector measurements outside the expected Field of View trajectory of dark photons. We found that the dark photon signal was consistent with zero, giving a new upper limit on the dark photon's kinetic mixing, $\chi$, with masses between 0.6 meV and 1.4 meV. These are the first constraints on dark photons as a dark matter candidate using an array of KIDs at millimeter wavelength.

\end{abstract}
\tableofcontents

\section{Introduction}

The search for dark matter is one of the greatest challenges in astroparticle and cosmology. One potential candidate for dark matter is the “dark photon” (DP). A non-thermal population of DP can be  produced during the early evolution of the Universe, with extensive literature on the different production mechanisms. For example, inflationary production distinguishes between longitudinal \cite{vectorDM, vectorDMEma, vectorDMSocha, vectorDMKolb}, and transverse modes \cite{vectorDMown, vectorDM1, vectorDM2}, the latter requiring an axion-like coupling between the inflaton and the DP. Transverse modes can be produced efficiently during inflation even when the DP is massless, contrary to the longitudinal ones. For our study case, a meV DP, the inflationary scale also plays an important role, as was reviewed in \cite{Beaufort:2023qpd}: transverse production with a relic abundance consistent with observations would require a lower inflationary scale than longitudinal production, implying a negligible prediction for the tensor-to-scalar ratio.
On the other hand, post-inflationary production can be efficient when the DP acquires its mass through a Dark Higgs (DH) mechanism. The DH field $h$ has the standard symmetry breaking potential
$V(h) = \frac{\lambda}{4} ( h^2 - v^2)^2 $,
where $v$ is the vacuum expectation value at the minimum. The masses of DP and DH are given by:
$m_X = g_D v \,,\; m_h = \sqrt{\lambda} v $,
where $g_D$ is the $U(1)_D$ coupling. 

Different scenarios depending on whether the symmetry is broken or not during inflation were considered in \cite{vectorDMRedi}. If the symmetry is already broken during inflation and the DH is heavy, i.e. $H_I< m_h$, $H_I$ being the inflationary Hubble parameter, the main contribution to the relic DM abundance comes anyway from inflationary production, with that from the Higgs decay being subdominant. Whereas in the case of a light DH, $m_h < H_I$, one can have, for example, resonant production of the DP due to the oscillations of the DH when inflation ends \cite{vectorDM3}; however, this would be suppressed for an meV DP that will privilege the scenario when the symmetry is broken after inflation and $H_I > v$. The production can be related to the dynamics of the dark Higgs \cite{vectorDMSato,vectorDMRedi}, or to the cosmic strings associated to the breaking of symmetry \cite{vectorDMCS1, vectorDMCS2,vectorDMCS3}. Cosmic strings will also give rise to a stochastic background of gravitational waves, which could be detectable by future gravity waves detectors. 

 Since the DP is the gauge boson of an Abelian U(1) symmetry, it can interact with the standard photon via a kinetic mixing \cite{Arias2012}. Assuming this is our only window towards the dark sector, and introducing the dimensionless parameter $\chi$ as the kinetic mixing, our starting point is the Lagrangian:
\begin{equation}
   {\cal L }= -\frac{1}{4}F_{\mu \nu}F^{\mu \nu}-\frac{1}{4}X_{\mu \nu}X^{\mu \nu}-\frac{\chi}{2}F_{\mu \nu}X^{\mu \nu}+\frac{1}{2}m_X^2X_\mu X^\mu + e J_\mu^{em} A^\mu \,,
\end{equation}
where $A^\mu$ is the standard photon with field strength $F^{\mu \nu}$, $X^{\mu}$ is the DP with field strength $X^{\mu \nu}$, and $J_\mu^{em}$ is the electromagnetic current.
The Dandelion experiment  was designed to detect these dark photons by exploiting their conversion into ordinary photons \cite{Fabbrichesi2021DarkPhoton} giving in addition a directional signature \cite{Beaufort2024Dandelion}. This paper presents the first results from our experiment, using data from a 1480-minute (24.7 hours) run.

The key idea behind our experiment is to use directionality to find the signal, as it was described in \cite{Beaufort:2023qpd}. As the Earth rotates, a signal from dark matter should move across our detector in a predictable path over 1480 minutes. This allows us to divide our detector pixels into two groups: a "Signal" group, where we expect to see the signal, and a "Background" group, which we use to measure only the noise.

The main difficulty is finding a very weak signal that is hidden in a large amount of background noise. This noise affects all our pixels simultaneously. We can take advantage of the signal and background noise properties to separate them. As discussed above the DP signal will be modulated and as so it will vary in shape and amplitude with time and from pixel to pixel. By contrast the background noise varies with time but it is expected to be correlated within pixels.


Our analysis confirms our previous estimations and shows the way to explore with a directional signature the DP at 1 meV, but did not find a positive signal, giving a new upper limit. This limit places new constraints on the dark photon's kinetic mixing, $\chi$. This paper describes our different analysis steps and presents our final result.

\section{Experimental Setup}
\begin{figure}[!ht]
    \centering
    
    \includegraphics[width=0.7\textwidth]{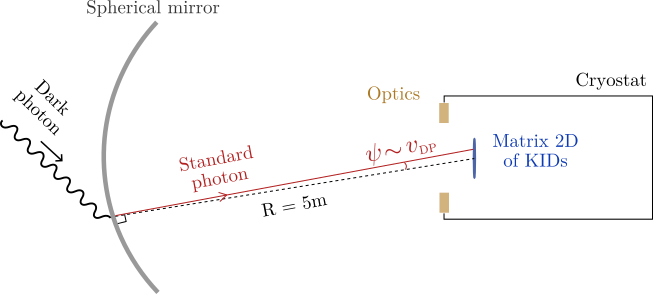}

    \caption{Schematic diagram of the experimental setup \cite{Beaufort2024Dandelion}, illustrating the operational DP-to-photon conversion process where a 5m-radius mirror emits signals toward the KISS-NIKA camera.}
    \label{fig:setup_and_optics}
\end{figure}

In this experiment, we use a spherical aluminum mirror with a diameter of 50 cm and a radius of curvature of 5 meters. This mirror serves as the conversion surface. If a dark photon from the galactic halo hits the mirror, it can be converted into a standard photon with an energy equal to the dark photon's mass, corresponding to a signal with a predicted wavelength of 1.24 mm.
These newly created photons are emitted nearly perpendicular to the mirror's surface. To detect those photons, we use an array of 221 Kinetic Inductance Detectors (KIDs) \cite{Zmuidzinas2012Review}, each with a sensitive surface area of $2.8 \times 2.8 \text{ mm}^2$. KIDs are high-quality microwave resonators that have a change in their electromagnetic properties in response to incoming radiation. The resonator is a superconductor which absorbs photons exceeding its gap energy, resulting in Cooper pairs breaking and thus, a change in the ratio of paired and unpaired (quasi-particles) charge carriers, which changes their resonant frequency, allowing the energy of an incident photon to be measured. The array is cooled to 150 mK inside the KISS-NIKA cryostat \cite{Monfardini2012,Mac2024} to minimize thermal noise (Fig. \ref{fig:setup_and_optics}). The system is optimized for frequencies around 250~GHz (corresponding to a wavelength of $\lambda \approx 1.20$~mm and a DP mass of $m_{\chi} \approx 1.03$~meV), aligning with the expected signature of 1~meV dark photon dark matter. The detector bandwidth ranges from 150~GHz ($\lambda \approx 2.00$~mm, $m_{\chi} \approx 0.62$~meV) to 350~GHz ($\lambda \approx 0.86$~mm, $m_{\chi} \approx 1.45$~meV); see Appendix~\ref{subsec:freq_to_mass}.

Between the mirror and the detector, a series of lenses guides the signal photons onto the detector array. The entire assembly, including the mirror and cryostat, is held by a rigid mechanical structure to ensure stable alignment, which was checked before and after the run with a laser.

The emission direction of the signal is set by the boundary conditions of electromagnetism at a conductive surface. At the surface of a good conductor like aluminum, the parallel (tangential) component of the electric field must be zero \cite{Horns2013WISPy}. 
When a dark photon converts at the mirror, the standard photon field produced must cancel the incoming tangential electric field in order to satisfy the boundary condition. This requirement forces the photon to be emitted at an angle shifted with respect to the perpendicular to the mirror’s surface. Due to the velocity of the dark photon in the galactic halo, there is a velocity component parallel to the mirror, $v_{||}$, which implies that the outgoing photon must acquire a corresponding momentum component parallel to the surface. This leads to a slight deviation from a perpendicular path, described by the angle $\psi$ \cite{Beaufort2024Dandelion}:
\begin{equation}
    \psi \approx \frac{v_{||}}{c} \approx 10^{-3} ~rad.
    \label{eq:emission_angle}
\end{equation}
In the KIDs matrix, we see that the distance between the spot of these signal photons and the center is equal to 5 mm; this offset represents the deviation of the signal from the perpendicular path. Critically, as the Earth moves and rotates, the direction of $v_{||}$ changes throughout the day as mentioned in \cite{Jaeckel2016Directional}. This causes the signal spot to trace a predictable path on the detector which represents our directional signature.
\begin{figure}[!ht]
    \centering
    \includegraphics[width=0.32\textwidth,height=5cm]{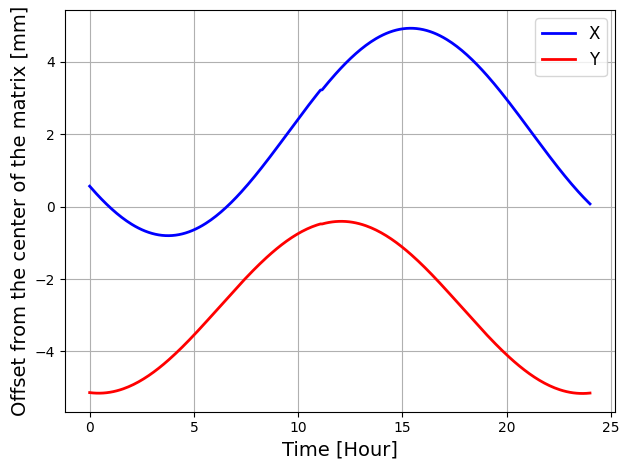}
    \hfill 
    \includegraphics[width=0.32\textwidth,height=5cm]{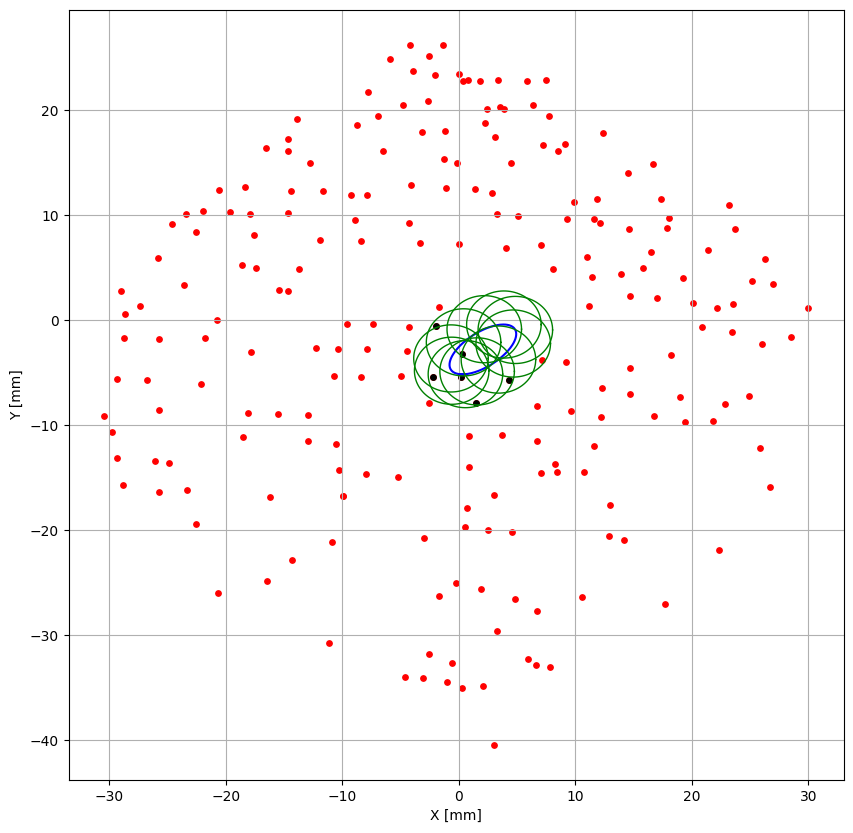}
    \includegraphics[width=0.32\textwidth,height=5cm]{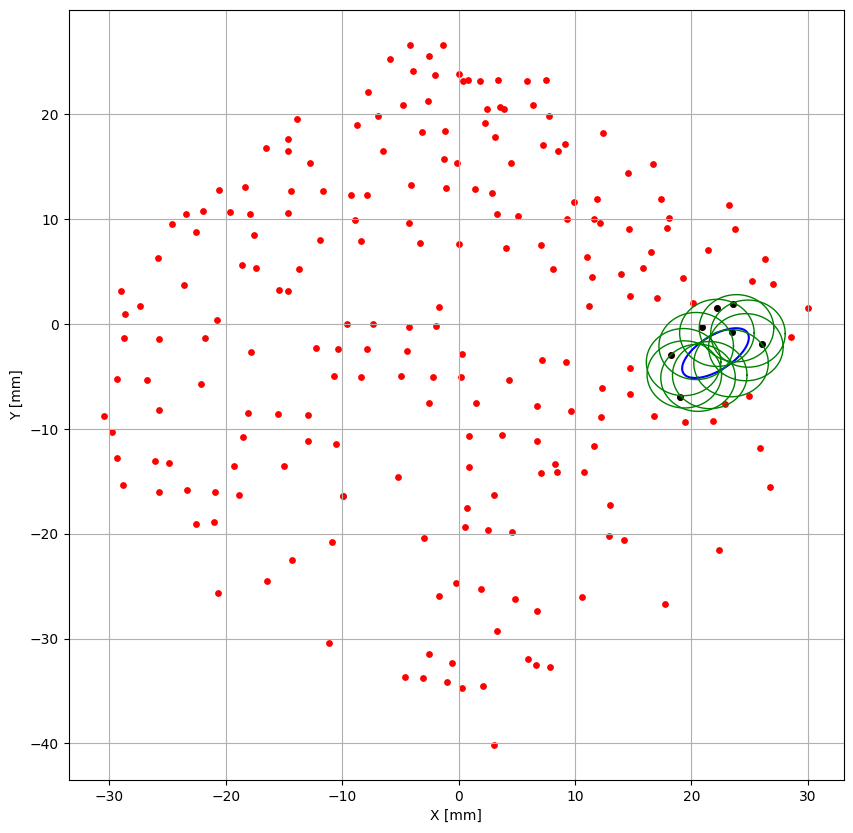}
    \caption{
    Spatial and temporal definition of the signal and background regions for the two-position analysis. The left panel shows the predicted 1480-minute run starting on January 11, 2024 temporal modulation of the signal's X and Y coordinates on the detector, as derived from the model in  \cite{Bozorgnia2011Modulation}. The other two panels project these complete daily trajectories onto the array of operational pixels for the first (frontal) and second (tilted) mirror positions. Based on proximity to these paths, pixels are classified into two critical sets: the 'Signal + Background' region (black dots), which is searched for the dark photon signature, and the 'Background-only' region (red dots).} 
    
    \label{fig:traj_posi} 
\end{figure}

% ...
During the 1480-minute measurement, we collected 148 data files of 10 minutes each. The mirror was switched between two positions, a "frontal" and a "tilted" alignment.  This change was made every second within each 10-minute data file. For the analysis, we use a mask to separate the data into two distinct time series, one for each mirror position. 
For the analysis, we concatenate the data from both mirror positions into a single, unified time series. This approach allows us to perform a global analysis on the entire 1480-minute dataset, effectively doubling the integration time compared to analyzing positions separately. 

\section{Signal modeling and reconstruction procedure}

For a given mirror position, using the NIKEL readout electronics \cite{Bourrion2016JInst,Bourrion2022JInst}, we measure the change in resonance frequency $\Delta f $ for each pixel, which is proportional to the input energy of the absorbed photon \cite{Bounmy2022,Fasano2021,Calvo2013}.

These frequency measurements (in Hz) for each pixel are then converted into temperature by multiplying by the pixel’s response (calibrated in K/Hz). The resulting time series for each pixel shows the evolution of the equivalent temperature shift corresponding to the power absorbed by that pixel during the observation period (1480 minutes on January 11, 2024). We have 148 data files, each lasting 10 minutes. To study the evolution of the absorbed temperature (or frequency) without being affected by cosmic glitches, we take the median over each 10-minute file, giving one point per file for a chosen position.  The absorbed power comes from several background sources (room temperature, thermal emission from the mirror and optical elements, stray light, etc.), in addition to the possible desired signal: standard photons produced from the conversion of dark photons. 
\subsection{Expected Dark Photon Signal}
Working in natural units where $c=1$, the expected power from dark photon conversion, measured at a position $(x, y)$ on the focal plane at a given time $t$, is described by the following equation:
\begin{equation}
    P(x, y, t) = \chi^2 \rho_{\text{CDM}} ~ \eta(E_\gamma) A_{\text{mirr}} \langle \cos^2 \alpha(t) \rangle_T I(x, y, t). 
    \label{eq:power_signal}
\end{equation}
\noindent $\rho_{\rm CDM}$ is the local dark matter density, $\eta(E_\gamma)$ is the detection efficiency, $A_{\rm mirr}$ is the surface area of the mirror, $I(x,y,t)$ the normalized signal spread on the detector, and $\alpha(t)$ the angle between the dark photon polarization and the tangent of the mirror, we use $\langle \cos^2 \alpha(t) \rangle_T \approx \frac{2}{3}$ \cite{Beaufort2024Dandelion}. For the purpose of our analysis, it is useful to separate this equation into two parts: an overall power amplitude and the spatial distribution $I(x,y,t)$. We can define the power amplitude, as:
\begin{equation}
    A_{\text{power}} =  \frac{2}{3} \chi^2 \rho_{\text{CDM}}~ \eta(E_\gamma)~ A_{\text{mirr}.}
    \label{eq:power_amplitude}
\end{equation}

\noindent The goal of the analysis presented in this paper is to estimate the kinetic mixing parameter using the temperature measured. This is done by first converting the temperature measurement of the dark photon signal into an equivalent power and  then we can deduce the coupling $\chi$. \\

\begin{figure}[!ht]
    \centering 
    \includegraphics[width=0.6\textwidth]{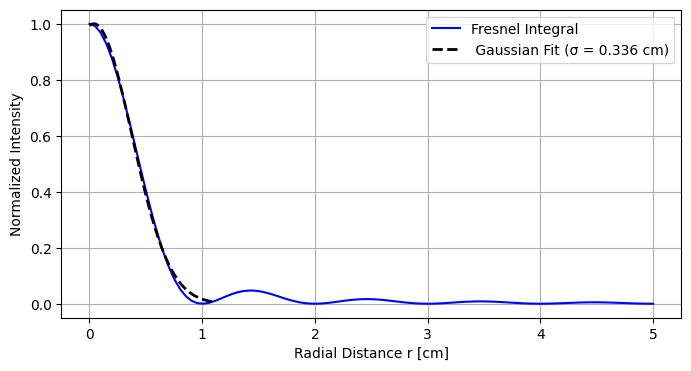} 
    \caption{Comparison between the change in intensity of the Fresnel diffraction function \cite{Beaufort2024Dandelion} and the two-dimensional Gaussian fit as a function of radial distance.} 
    \label{fig:mir}
\end{figure}

The function $I(x,y,t)$ given by Eq. (\ref{eq:power_signal}) represents the Fresnel diffraction intensity pattern  produced by the finite aperture of the spherical mirror on the detector plane. 
Its shape is set by diffraction, while the spot center $(x_c(t), y_c(t))$ (one of the points of the trajectory) evolves over 1480~minutes due to the Earth's motion, but its shape at any given instant is determined by this diffraction pattern (Fig. \ref{fig:mir}).

Since the signal power is directly converted into a measurable temperature change in our KIDs, the temperature signature we search for must follow this same spatial distribution. That is, we expect the excess temperature on the detector, $T_{\text{signal}}$, to be directly proportional to the intensity profile $I(x,y,t)$:
\begin{equation}
    T_{\text{signal}}(x, y, t) \propto P(x, y, t) \propto I(x,y,t)
\end{equation}

While the exact mathematical form of a Fresnel diffraction pattern is complex, its central lobe—where most of the signal power is concentrated—can be accurately approximated by a two-dimensional Gaussian function. To precisely define our signal model, we performed a fit of the theoretical Fresnel diffraction profile  using a normalized two-dimensional Gaussian function, as illustrated in Fig.~\ref{fig:mir}. The fit gives a standard deviation of $\sigma_D = 3.2 \pm 0.04 ~ mm$. For the purposes of our analysis, we adopt this empirically determined value as a fixed parameter for the signal width (diffraction). Modeling the signal profile as a Gaussian function  offers two key advantages: it is analytically straightforward and computationally efficient for the final fitting procedure. %

\subsection{Background contribution}
\begin{figure}[!ht]
    \centering

    \includegraphics[width=0.32\textwidth]{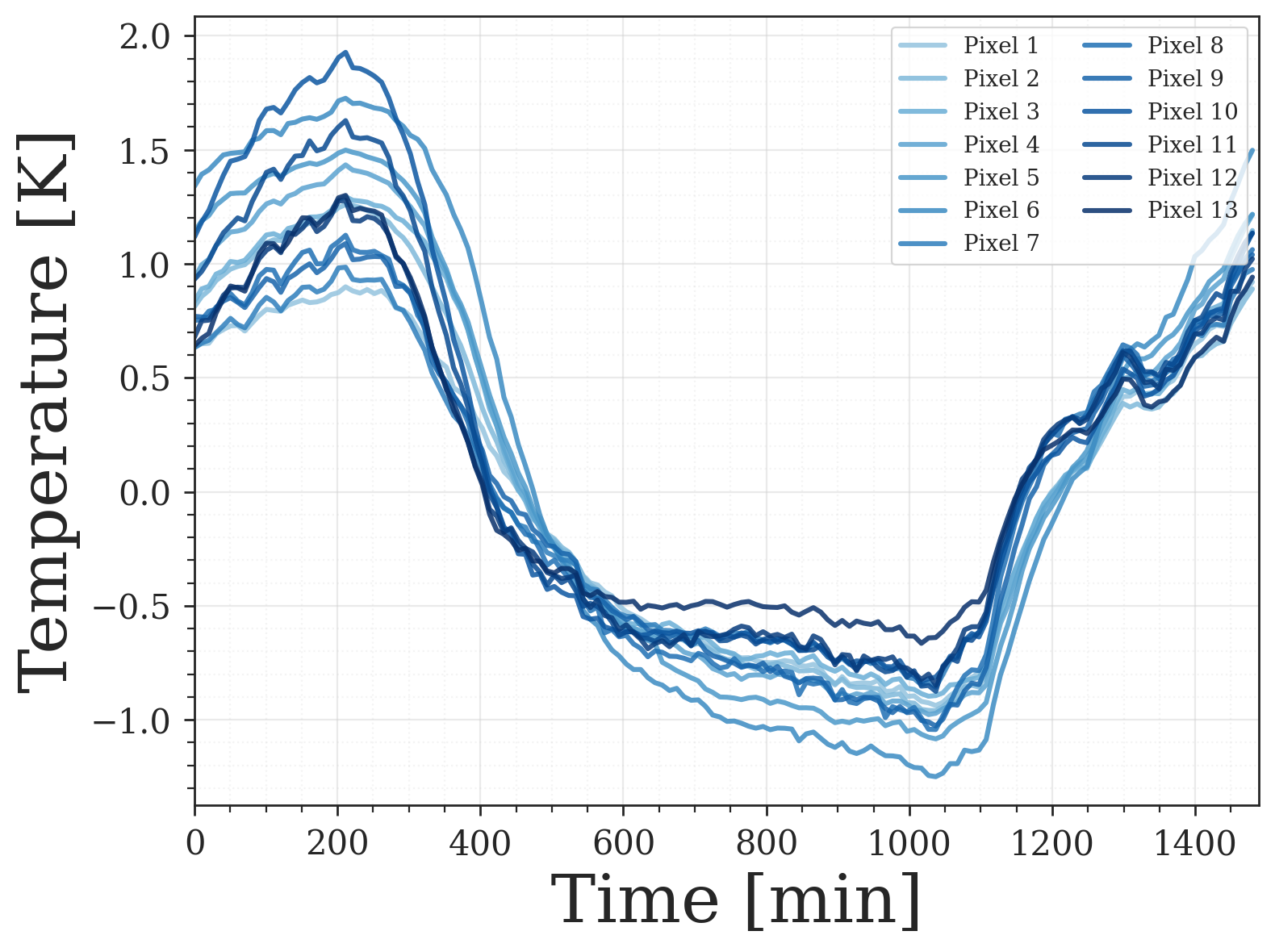}
    \includegraphics[width=0.32\textwidth]{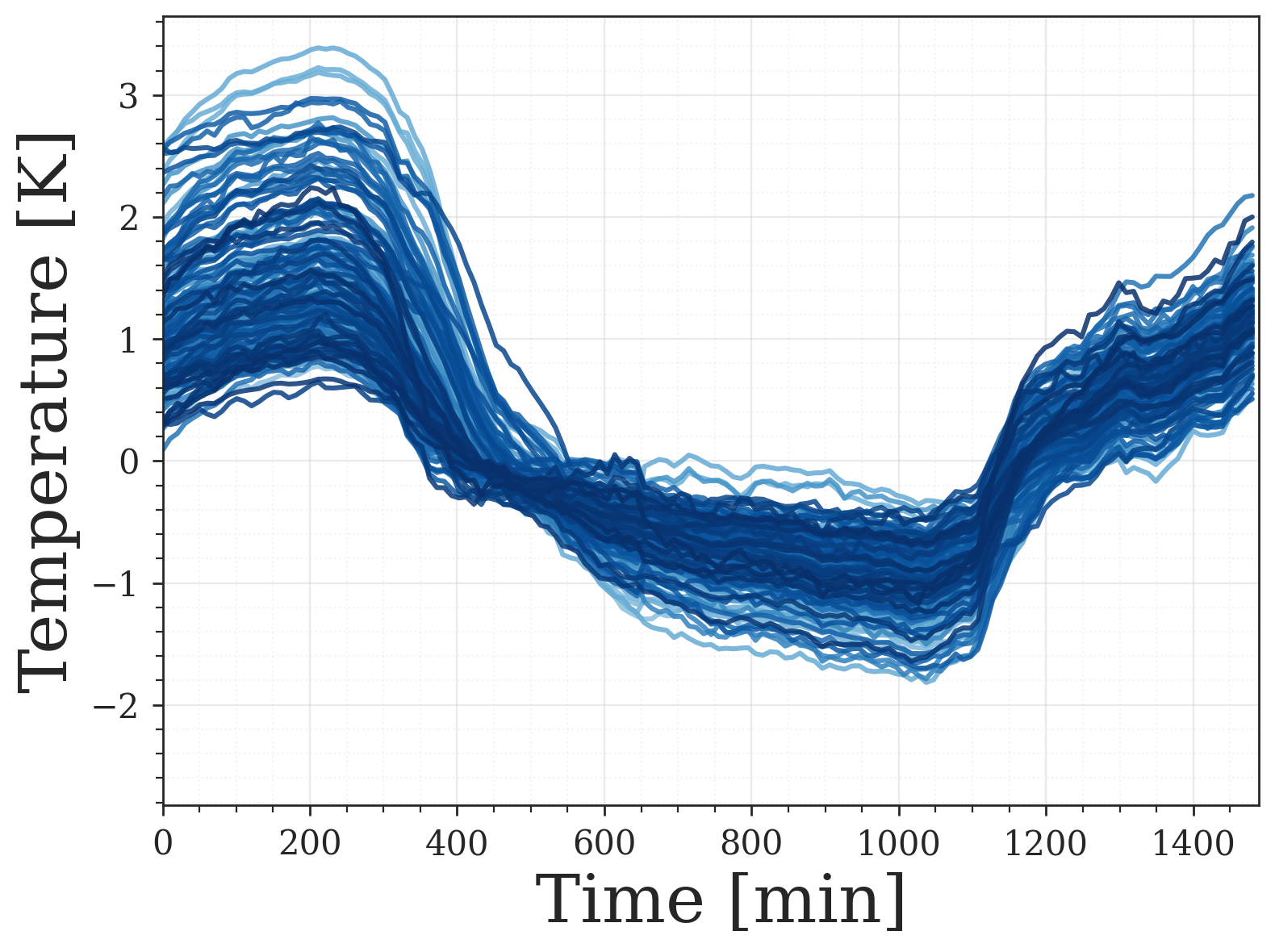}
    \includegraphics[width=0.32\textwidth]{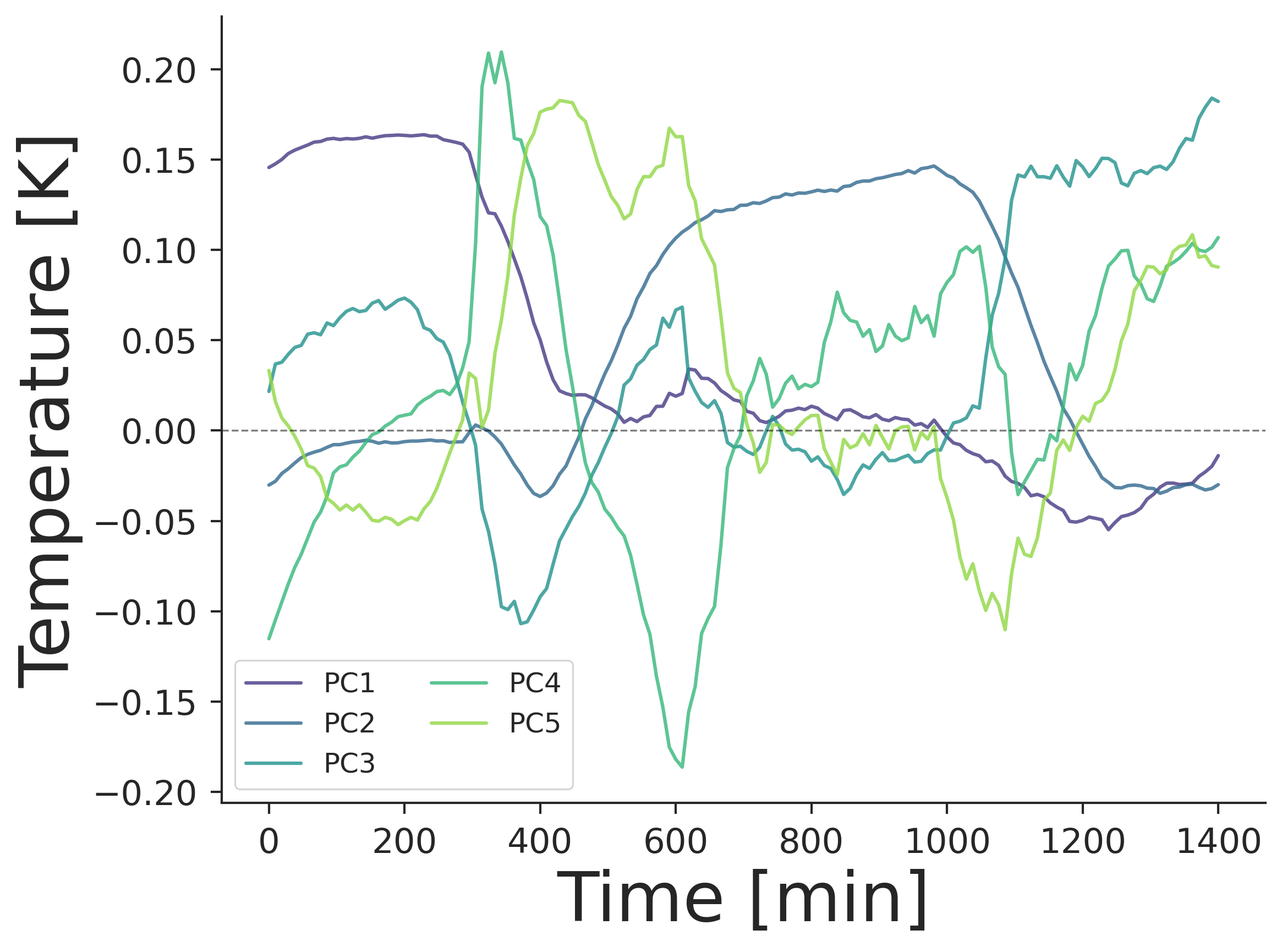}
    \caption{The left and middle panels show the raw temperature time series for a subset of 'Signal' pixels and the full set of 'Background' pixels, respectively. Both are dominated by large, correlated thermal fluctuations. To model this common-mode noise, we apply Principal Component Analysis (PCA) to the concatenated background data. The right panel displays the first five (out of 10 used) resulting principal components, which represent the dominant,  independent modes of temporal variation. Note that these principal components have been normalized for visualization purposes, which is why their vertical scale does not visually correspond to the large, unscaled fluctuations seen in the raw data of the left and middle panels. These components are used to construct the noise model for subtraction. }
    \label{fig:tempevo}
\end{figure}

The evolution of temperature (Fig.~\ref{fig:tempevo}) shows a correlation between pixels caused by uniform thermal background noise that affects the entire array. The main contribution arises from the thermal emission of the mirror and the surrounding structure at room temperature, which continuously radiate in the millimeter range. Additional power is emitted and reflected by the optical lenses located at different temperature stages (300 K, 4 K, and 150 mK). Even though the coldest lenses emit very little, the warmer ones still send a noticeable flux toward the detectors. A smaller contribution comes from electronic and readout noise, as well as residual stray light entering the cryostat through imperfect shielding or reflections. All these sources vary slowly with time because of temperature changes in the cryostat, or fluctuations in the electronics, and because the pixels see the same optical Field of View (FOV), their signals show a strong temporal correlation. And the most important fluctuations come from the mirror temperature and can lead to variations of the order of a Kelvin.

\subsection{Two-region model}

\begin{figure}[!ht]
    \centering
   
    \includegraphics[width=0.48\textwidth]{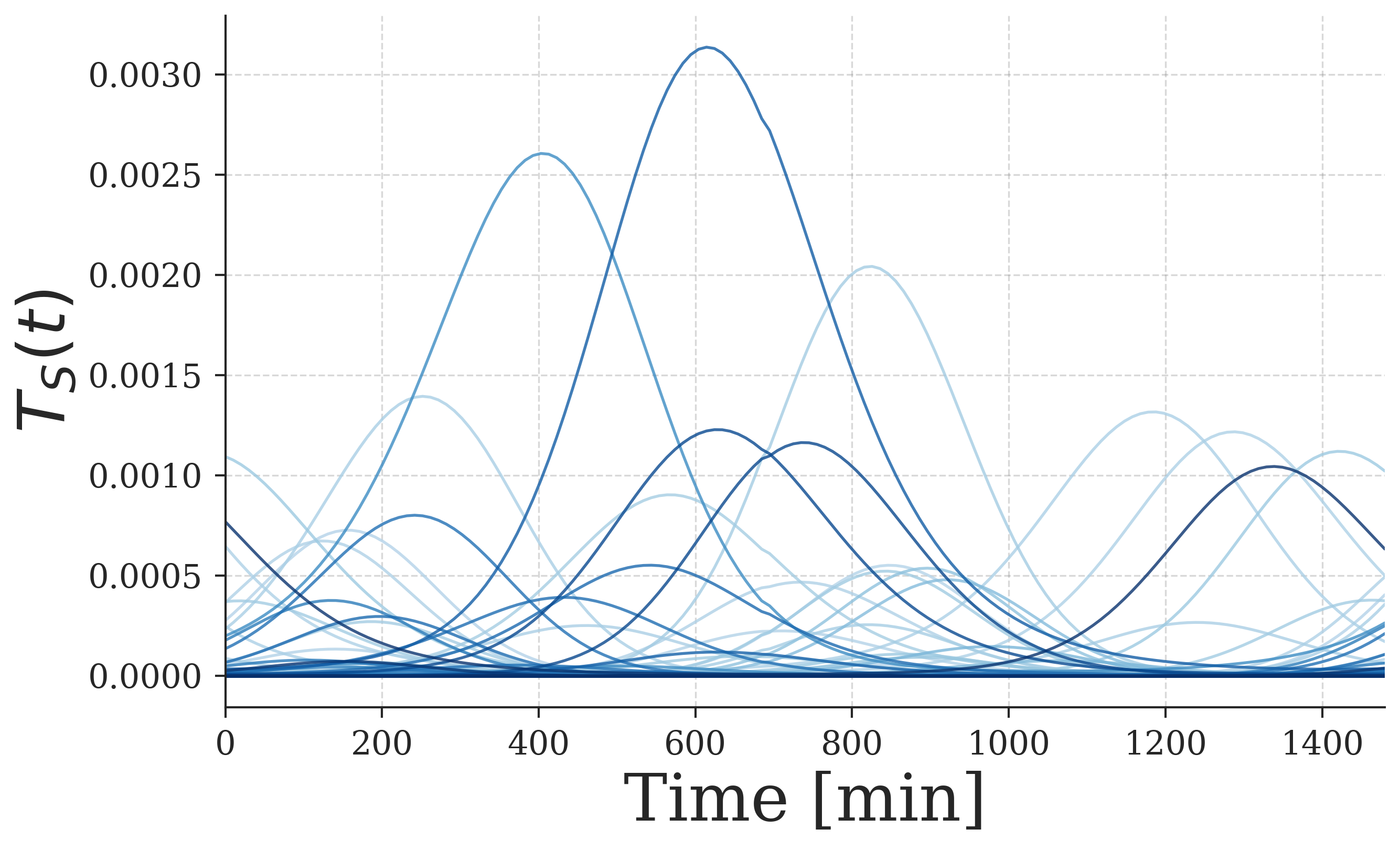}
    \includegraphics[width=0.48\textwidth]{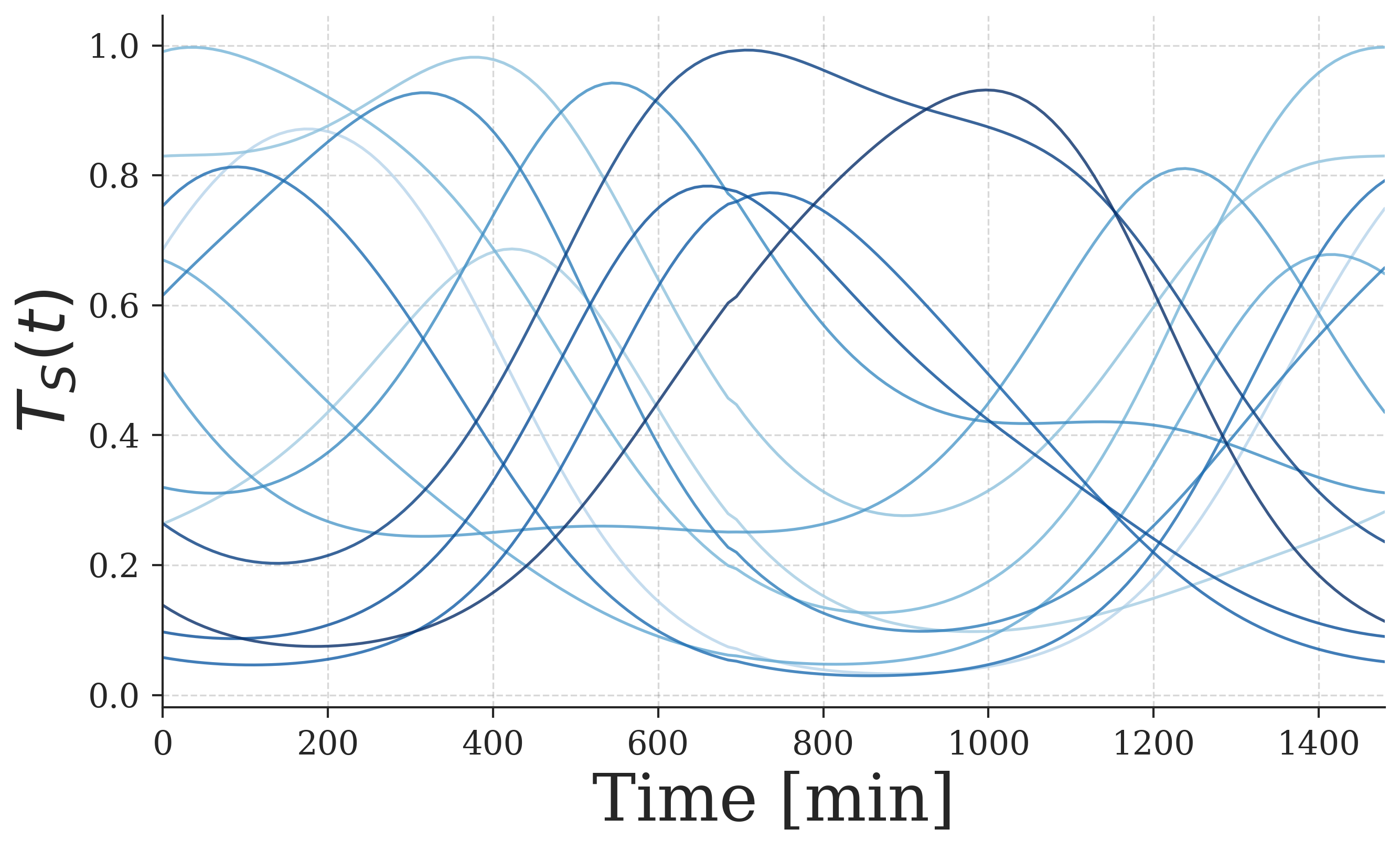}
    \caption{ This figure illustrates the modeled temporal evolution of the Gaussian signal component for pixels in both the \texttt{Signal} and \texttt{Background} regions. 
    \textbf{Right Panel:} The normalized signal templates for pixels located within the \texttt{Signal + Background} region. As the Gaussian signal spot moves along its daily trajectory, each pixel is illuminated at a distinct time, creating the unique temporal signature that is the target of our search. 
    \textbf{Left Panel:}  The corresponding signal component for a set of pixels in the \texttt{Background-only} region. The amplitude is three orders of magnitude lower, remaining negligible throughout the observation. This quantitatively justifies their use for a clean extraction of common-mode noise.}
    \label{fig:pixel_selection}
\end{figure}

Our analysis strategy relies on separating the signal from the background using directionality (the signal trajectory). We divide the pixels of the array into two regions (Fig.~\ref{fig:pixel_selection}):  
\begin{enumerate}
    \item \textbf{Background-only region}: pixels that are never on the signal trajectory and are therefore unaffected by the diffraction spot. These pixels only measure the background.
    \item \textbf{Background + Signal region}: pixels located along the expected signal trajectory.
\end{enumerate}


To isolate the potential dark photon signal from the dominant background noise, we construct a model for the temperature measured by each pixel over time. Our model assumes that the data from any given pixel is a sum of a background component and a potential dark photon induced signal component  (Eq. \ref{eq:bkgsign}). 
\begin{equation}
    T^{traj}_{k}(t) =b_{k} + \sum_{i=1}^{N_{sys}} a_{k,i} \cdot T^{\mathrm{syst
    }}_i(t) + T_{\text{S}} \cdot 
    \exp\left(-\frac{(x(t)-x_{k})^2 + (y(t)-y_{k})^2}{2\sigma_{D}^2}\right).
    \label{eq:bkgsign}
\end{equation}
The background component, which affects all pixels, is described as a linear combination of time-dependent noise templates, $T^{\mathrm{syst}}_i(t)$, $T_{S}$ is the dark photon associated signal amplitude in Kelvin, $(x(t),y(t))$ is the signal trajectory, and $(x_{\text{k}}, y_{\text{k}})$ represents the position of the pixel in the array. 
The coefficients $b_{k}$ (offset) and $a_{k,i}$ (slopes) are unique to each pixel "k" and define its specific response to these shared background fluctuations.
The signal component, $S(x_{k}(t), y_{k}(t))$, is different. We only expect it to appear in the pixels that fall along the predicted signal path at a given time t.

The pixels that are far from the trajectory path, and are thus unaffected by the dark photon signal, provide a clean measurement of the dominant common-mode noise, which is then modeled and subtracted from the entire dataset to enable a sensitive search for the faint, directionally varying signal.

Figure \ref{fig:pixel_selection} provides a quantitative justification for dividing the pixels into two regions. The simulated Gaussian profile shows that the farther a pixel is from the dark photon trajectory, the smaller its expected signal amplitude becomes. In the background region, the dark photon contribution is therefore negligible. This comparison confirms that these pixels provide a clean measurement of the common-mode noise.

\subsection*{Determination of the background templates}

The observed background has multiple contributions, including the thermal 
emission of the mirror at 300~K, the emissivity of the three lenses (at 300~K, 
4~K, and 150~mK), electronic noise, and stray light. Estimating a 
single function that accurately describes all these contributions is particularly 
challenging.

The difficulty mainly arises from the uncertain determination of 
the absorption and emission coefficients of the optical system, which depend 
directly on the material properties. These properties themselves vary strongly 
with temperature, further complicating the precise evaluation of these coefficients  and, consequently, the overall modeling of the background.

To solve this problem, we apply Principal Component Analysis (PCA) \cite{Smith2002PCA} to the pixel data from the  ``Background'' region for each mirror position. PCA  is a decomposition that identifies the common temporal variation modes (the principal components, $T^{\mathrm{syst}}_i(t)$) which capture the majority of the correlated background noise. Modeling the background noise using PCA is a powerful technique to extract  patterns of common variation (Fig. \ref{fig:tempevo}) from a complex, correlated dataset. In our case,  the data consist of the temporal time series of pixels in the ``Background''  region, where we are confident that only noise is present. 
To ensure the stability of our analysis, we verified that performing two parallel and independent PCA analyses yielded equivalent background models and final exclusion limits across both mirror positions. While the two-position modulation allows us to observe background fluctuations every second, the decision to use a unified PCA approach was made to effectively capture common-mode noise that was otherwise difficult to decouple between the two individual positions. By concatenating these background-only intervals to train the PCA, we ensure that the resulting noise model is fully representative of the conditions during the entire acquisition period.

\begin{figure}[!ht]
    \centering

    \includegraphics[width=0.7\textwidth]{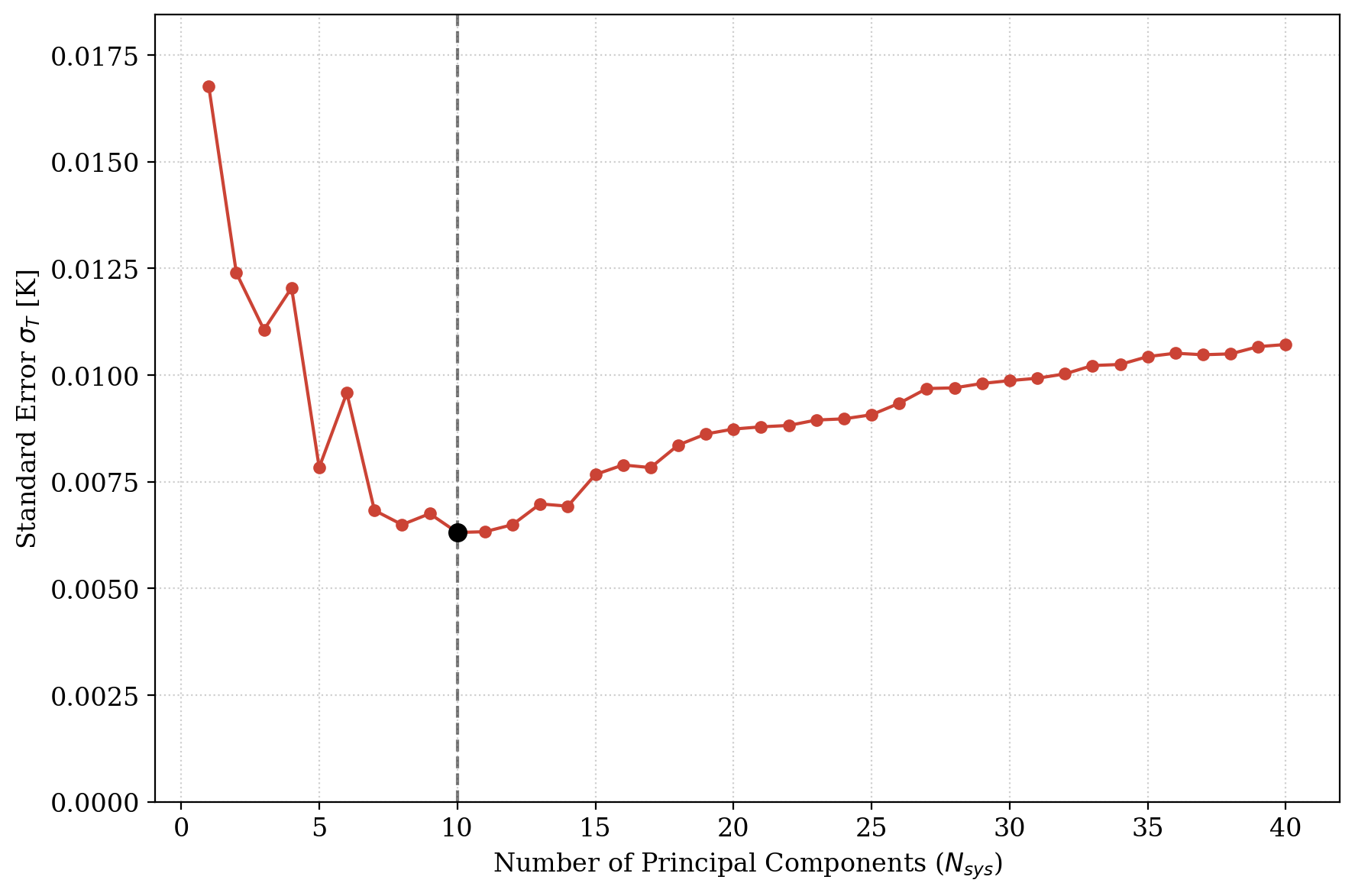}
    \caption{Uncertainty of the signal amplitude ($\sigma_T$) plotted as a function of the number of principal components ($N_{\text{sys}}$) used in the noise model. To ensure a consistent analysis, a single optimal value of \textbf{$N_{\text{sys}} = 10$} is  selected where the uncertainty reaches its minimum (red curve), providing a robust basis for our noise subtraction across the concatenated dataset.}
    \label{fig:cp_selection}
\end{figure}

We select 190 pixels in the background-only region. We concatenate their measurements from both mirror positions to form a single dataset and decompose the signal into independent components using a Principal Component Analysis (PCA). Note that these independent components will include both correlated contributions across all detectors and the noisy contributions. Therefore, 
the choice of the number of Principal Components ($N_{\text{sys}}$) used to model the background noise (correlated between detectors) is a critical step in the analysis. This choice is guided by the goal of minimizing the final uncertainty on the signal amplitude ($\sigma_T$). Although the noise characteristics can differ slightly between the two datasets, leading to slightly different optimal PC counts, we chose a single number of components for consistency. As shown in Figure~\ref{fig:cp_selection}, we selected \textbf{$N_{\text{sys}} = 10$} for both mirror positions. This value represents a robust choice, located in the region where the signal uncertainty is  at its minimum, ensuring a balanced and effective noise subtraction.

\subsection{Final analysis}
To fit our model to the data, we use a linear regression framework based on Maximum Likelihood Estimation (MLE). For each pixel, we construct a likelihood function based on the assumption of Gaussian noise.

By maximizing this function, we find the best-fit values for all the model parameters: the offset, the contribution of each Principal Component $T^{\mathrm{syst}}_i(t)$, and the amplitude of the signal $T_{\text{S}}$. This provides a robust noise model for each pixel, which can then be subtracted to search for a significant signal. For pixels near the expected trajectory, the signal component, proportional to the Gaussian function in Eq.(\ref{eq:bkgsign}), is included in the fit, allowing us to directly constrain its amplitude.

\begin{figure}[!ht]
    \centering

    \includegraphics[width=\textwidth]{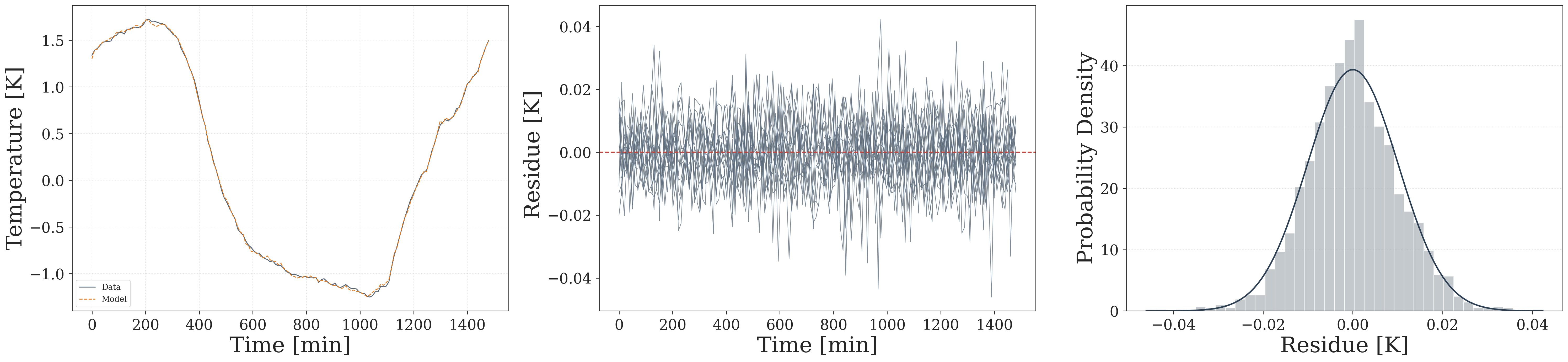}
   
      \caption{Performance of the noise subtraction procedure. Left: Time evolution of the raw temperature data (solid line) compared to the model (dashed line) for a representative pixel, showing the model's ability to accurately track large-scale thermal fluctuations. Center: Resulting temperature residuals for all pixels in the signal region after model subtraction; the fluctuations are centered around zero with a significantly reduced amplitude compared to the raw data. Right: Histogram of the temperature residuals from all pixels, fitted with a Gaussian profile to demonstrate that the remaining noise is stochastic.}

        \label{fig:resi}
\end{figure}

The fit (Fig. \ref{fig:resi}) shows good agreement between the observed data and the background noise model, validating the PCA approach. We find $T_{\text{signal}} = 0  \pm 6~\text{mK}$. No statistically significant $T_{\text{signal}}$ amplitude was observed. Therefore, we estimate the upper limit of the temperature amplitude as $\Delta T_{\text{upper-limit}} = 12~\text{mK}$. Translating this temperature limit into constraints on the power allows us to compute the exclusion curve for the kinetic mixing parameter.
\section{Exclusion Curve}
Analysis of the data revealed no statistically significant signals above background fluctuations. We therefore proceeded to set an upper limit on the physical parameters of the dark photon model. This process breaks down into three key steps: determining the limit on the temperature amplitude, then converting it into a power limit, and finally calculating the constraint on the kinetic mixing coupling,  $\chi$. In the absence of detection, we set an upper limit on signal amplitude. Linear regression enables us to set a 95\% confidence limit on temperature amplitude: $\Delta T < \SI{12}{m\kelvin}$.

The next step is to translate this temperature limit into a limit on the power incident on the detector. The relationship between power and temperature variation $\Delta T$, for a blackbody detector, is described by Stefan-Boltzmann's law, integrated over the bandwidth $ H(\nu)$, $A\Omega$ is the throughput of each pixel, $\zeta(\nu)$ is the efficiency of the optical system composed of the three lenses (with $\langle \zeta(\nu) \rangle = 0.6$), and $\langle \eta (\nu)\rangle  = \langle \zeta(\nu)\cdot H(\nu) \rangle =0.3$ represents the detection efficiency seen in equation \ref{eq:power_amplitude}. The power amplitude is given by:
\begin{equation}
 A_{\text{power}} = A\Omega \int_{\nu_{min}}^{\nu_{max}} \zeta(\nu) H(\nu) \Delta B(\Delta T, \nu)  d\nu.
\label{eq:power_conv}
\end{equation}

In our operating regime, where the total system temperature is high relative to the photon energy we want to measure ($k_B T_{\text{tot}} \gg h \nu$), we can use the Rayleigh-Jeans approximation, so we have $\Delta B = B( T_{background} + \Delta T, \nu) -B( T_{background}, \nu)$. This condition is largely satisfied, especially for the 300~K lens, where the thermal energy is around 25 times greater than the energy of the 244~GHz photon. By calculating the integral, taking into account the measured bandwidth and the limit $\Delta T_{\text{lim}}$, we can convert the temperature limit into a detectable power limit. This limit on power is then converted into a limit on kinetic mixing coupling $\chi$  (Eq. \ref{eq:power_amplitude}). The final expression for the coupling is:
\begin{equation}
\chi = \sqrt{\frac{ A_{\text{power}}}{\rho_{\text{CDM}} \cdot A_{\text{mirror}} \cdot \frac{2 }{3} \cdot c \cdot \langle \eta (\nu)\rangle }}  ~~.
\end{equation}

The analysis was performed for two different mirror positions (m1 and m2) assuming a local dark matter density $\rho_{CDM} = 0.44 \frac{GeV}{cm^3}$ \cite{Soding2025} and $A_{mirror} = 1963 ~ \text{cm}^{2}$. The result obtained by combining both position datasets, for frequencies between 150 GHz and 350 GHz corresponding to masses between 0.6 meV and 1.4 meV, is  $\chi_{} <  8.7 \times 10^{-10}$, which corresponds to a 95\% confidence level upper limit.
\begin{figure}[!ht]
    \centering
    \includegraphics[width=0.6\textwidth]{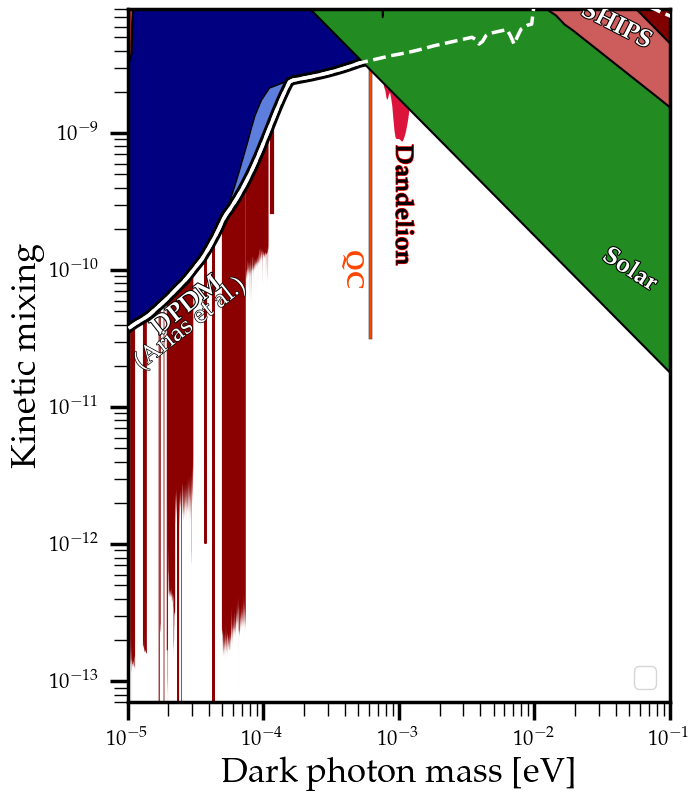} 

    \caption{ Combined 95\% confidence level  exclusion limit on the kinetic mixing parameter, $\chi$, as a function of the dark photon mass, \(m_{X}\). 
    This limit is derived under the assumption that dark photons constitute the entirety of the local dark matter density (\(\rho_{\text{CDM}} \approx 0.44 \, \text{GeV/cm}^3\)).  
    The solid red band represents the new constraint set by this work, the Dandelion experiment, with only 24.7 hours measurement, which probes the mass range from 0.6\,meV to 1.4\,meV. 
    Existing limits from other experiments are shown as shaded regions for context. Figure adapted from the AxionLimits  Github of C.~O'Hare~\cite{cajohare_axionlimits}.}
    \label{fig:exclusion}
\end{figure}

\section{Conclusion}
In summary, the Dandelion experiment has carried out its first directional search for dark photon dark matter using a 1480-minute dataset. Despite the presence of strong correlated background noise, our Principal Component Analysis method enabled an efficient subtraction procedure, providing sensitivity to potential signals along the predicted daily trajectory. No significant excess was observed, allowing us to establish new exclusion limits on the kinetic mixing  $\chi$ in the mass range 0.6–1.4 meV. These results represent the first constraints from the Dandelion  experiment and demonstrate the feasibility and power of our approach.

\section*{Acknowledgments}

The authors acknowledge funding from the French Programme d’investissements d’avenir through the Enigmass Labex. MBG acknowledges partial support by PID2022-140831NB-I00 funded by MICIU/AEI/10.13039/501100011033 and
FEDER,UE, and FCT CERN grant 10.54499/2024.00\\252.CERN.  

\section*{Appendix}
\appendix
\section{Frequency Bandwidth}

\label{subsec:freq_to_mass}

The primary goal of the Dandelion experiment is to detect dark photon dark matter ($X$) from the galactic halo. Particles in the halo are non-relativistic, with a typical velocity dispersion of $v \approx 10^{-3}c$. Consequently, the total energy of a dark photon is overwhelmingly dominated by its rest mass energy, with its kinetic energy being negligible.

The detection mechanism relies on the conversion of a dark photon into a standard photon ($\gamma$) at the surface of the conductive mirror. According to the law of energy conservation, the energy of the standard photon produced, $E_{\gamma}$, must be equal to the energy of the incident dark photon. Therefore:
\begin{equation}
    E_{\gamma} \approx E_{X} \approx m_{X}c^2 \implies \quad m_{X} = \frac{h\nu}{c^2}
\end{equation}


This allows us to translate the frequency range of our detector's sensitivity into a specific mass range for our dark photon search. For instance, the center frequency of our filter at $\nu = \SI{250}{\giga\hertz}$ corresponds to a dark photon mass of $\SI{1}{\milli\electronvolt\per\cancel{c^2}}$.

\begin{figure}[!ht]
    \centering
    \includegraphics[width=0.5\textwidth]{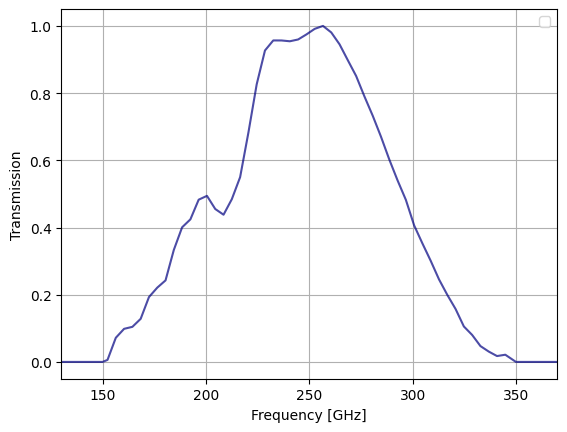} 
    \caption{Measured Transmission Spectrum of the Band-Pass Filter. The plot shows the experimental frequency response of the filter from 150 GHz to 350 GHz. The filter exhibits a center bandwidth frequency of 250 GHz with a peak transmission of nearly 100$\%$.}
    \label{fig:bandwi}
\end{figure}

The measured band-pass filter response shown in Fig.~\ref{fig:bandwi} directly defines the experimental bandwidth used in our analysis. The system is optimized to observe at a central frequency of $250\,\mathrm{GHz}$ with a bandwidth of around $0.2$. The bandwidth is defined as
\[
\frac{\Delta \nu}{\nu_0} = \frac{\mathrm{FWHM}}{\nu_0},
\]
where $\mathrm{FWHM}$ is the full width at half maximum of the spectrum and $\nu_0$ is the central frequency.

All limits presented in the exclusion curve are therefore derived assuming this effective frequency range of the bandwidth, which sets the sensitivity of the Dandelion experiment to dark photon masses between 0.6 meV and 1.4 meV.

\section{Rayleigh-Jeans Approximation}
\label{sec:appendix_rj}
In our analysis, we work directly within the Rayleigh-Jeans approximation, valid in the regime $k_B T \gg h\nu$. In this limit, the spectral radiance $B_{RJ}$ is given by:
\begin{equation}
    B_{RJ}(\nu, T) \approx \frac{2\nu^2 k_B T}{c^2}.
    \label{eq:rj_law}
\end{equation}
This result shows a direct linear relationship between spectral radiance and temperature.

In our experiment, we search for a very small signal temperature modulation, $\Delta T$, superimposed on a large, quasi-stable background temperature, $T_{\text{background}}$. The power associated with the signal is therefore proportional to the difference in radiance:
\begin{equation}
    \Delta B_{RJ}(\nu) \approx \frac{2\nu^2 k_B}{c^2} (T_{\text{background}} + \Delta T) - \frac{2\nu^2 k_B}{c^2} T_{\text{background}} = \frac{2\nu^2 k_B}{c^2} \Delta T.
\end{equation}

The validity of this approximation in our setup can be verified by comparing the energy scales. Our signal photons have frequencies centered around $\nu \approx \SI{244}{\giga\hertz}$, corresponding to an energy of $E_{\gamma} = h\nu \approx \SI{1}{\milli\electronvolt}$. The dominant thermal background is from optical elements at room temperature, $T_{\text{background}} \approx \SI{300}{\kelvin}$, which corresponds to a thermal energy of $k_B T_{\text{background}} \approx \SI{25.84}{\milli\electronvolt}$.

Since $k_B T_{\text{background}} \gg h\nu$ (by a factor of approximately 25), the Rayleigh-Jeans condition is well satisfied. This linear relationship justifies the procedure described in Section~4, where a limit on the measured temperature amplitude, $\Delta T_{\text{limit}}$, is directly converted into a limit on incident signal power, $A_{\text{power}}$.

\section{Synchronous Modulation Technique}

\begin{figure}[!ht]
    \centering
    \includegraphics[width=0.8\textwidth]{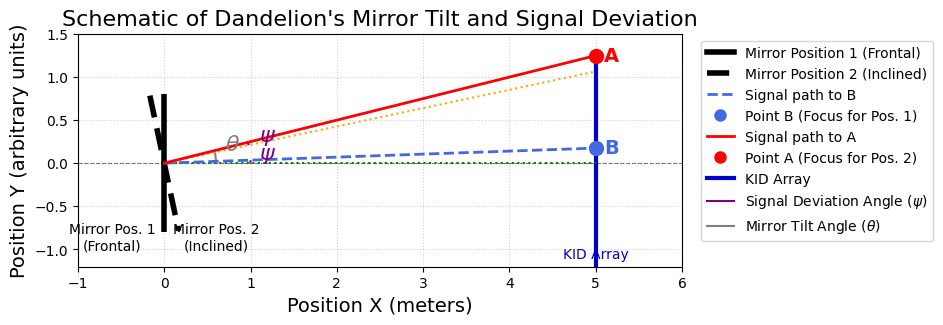} 
    \caption{Schematic illustration of the two-position mirror modulation strategy. The primary mirror is alternated between a "frontal" alignment (Position 1) and a "tilted" alignment (Position 2). The controlled tilt of the mirror by an angle~$\theta$ displaces the signal's focal point on the KID array from point B to point A. This schematic distinguishes the instrumental tilt~$\theta$ from the intrinsic signal deviation angle~$\psi$, which represents the emission angle of photons relative to the mirror's normal.}
    \label{fig:schematic_two_pos}
\end{figure}

This appendix provides further technical details on the two-position mirror modulation strategy, which was implemented to provide two independent datasets for cross-validation. The principle is schematically illustrated in Figure~\ref{fig:schematic_two_pos}.

The primary mirror was alternated between a "frontal" position (Position 1), where the optical axis is aligned with the detector center, and a "tilted" position (Position 2). It is important to note that the signal itself, from dark photon conversion, is inherently emitted at a small deviation angle, denoted by~$\psi$, which represents the exit angle of the signal with respect to the perpendicular to the mirror's surface. The controlled tilt angle of the mirror, $\theta$, was then applied in Position 2 to displace the focal point of this already-deviated signal from point B to point A on the detector plane. This displacement ensures that the two signal trajectories illuminate different sets of pixels, as shown in Figure~\ref{fig:traj_posi}.

This technique allowed us to perform two parallel and independent analyses. By using the results from both positions, we could verify that any residual background did not mimic a dark photon signal, thus significantly increasing the robustness of our final exclusion limit.

\bibliographystyle{JHEP}
\bibliography{kha} 

@article{Beaufort2024Dandelion,
      author        = {Beaufort, C. and Bastero-Gil, M. and Catalano, A. and Erfani-Harami, D.-S. and Guillaudin, O. and Macias-Perez, J. and Santos, D. and Savorgnano, S. and Vezzu, F.},
      title         = {{Directional detection of meV dark photons with Dandelion}},
      journal       = {arXiv:2401.07724},
      archivePrefix = {arXiv},
      eprint        = {2401.07724},
      primaryClass  = {hep-ph},
      year          = {2024},
      month         = {jan}
}

@article{Horns2013WISPy,
      author        = {Horns, D. and Jaeckel, J. and Lindner, A. and Lobanov, A. and Redondo, J. and Ringwald, A.},
      title         = {{Searching for WISPy Cold Dark Matter with a Dish Antenna}},
      journal       = {JCAP},
      volume        = {2013},
      number        = {04},
      pages         = {016},
      year          = {2013},
      doi           = {10.1088/1475-7516/2013/04/016},
      eprint        = {1212.2970},
      archivePrefix = {arXiv},
      primaryClass  = {hep-ph}
}

@article{Jaeckel2016Directional,
      author        = {Jaeckel, Joerg and Knirck, Stefan},
      title         = {{Directional Resolution of Dish Antenna Experiments to Search for WISPy Dark Matter}},
      journal       = {JCAP},
      volume        = {2016},
      number        = {01},
      pages         = {005},
      year          = {2016},
      doi           = {10.1088/1475-7516/2016/01/005},
      eprint        = {1509.00371},
      archivePrefix = {arXiv},
      primaryClass  = {hep-ph}
}

@article{Fabbrichesi2021DarkPhoton,
      author        = {Fabbrichesi, Marco and Gabrielli, Emidio and Lanfranchi, Gaia},
      title         = {{The Dark Photon}},
      journal       = {Annu. Rev. Nucl. Part. Sci.},
      volume        = {71},
      pages         = {401--427},
      year          = {2021},
      doi           = {10.1146/annurev-nucl-102419-041432},
      eprint        = {2005.01515},
      archivePrefix = {arXiv},
      primaryClass  = {hep-ph}
}

@article{Soding2025,
    author = {S\"{o}ding, Laurin and Bartel, Ruben L.},
    title = "{Local dark matter density from Gaia DR3 K-dwarfs using Gaussian processes}",
    journal = {Monthly Notices of the Royal Astronomical Society},
    volume = {542},
    number = {3},
    pages = {2987-2997},
    year = {2025},
    month = {08},
    note = {\url{https://doi.org/10.1093/mnras/staf1391}}
}

@article{Zmuidzinas2012Review,
      author        = {Zmuidzinas, Jonas},
      title         = {{Superconducting Microresonators: Physics and Applications}},
      journal       = {Annu. Rev. Condens. Matter Phys.},
      volume        = {3},
      pages         = {169--214},
      year          = {2012},
      doi           = {10.1146/annurev-conmatphys-020911-125022},
      eprint        = {1204.5334},
      archivePrefix = {arXiv},
      primaryClass  = {cond-mat.supr-con}
}

@article{Bozorgnia2011Modulation,
      author        = {Bozorgnia, N. and Gelmini, G. B. and Gondolo, P.},
      title         = {{Daily modulation due to channeling in direct dark matter crystalline detectors}},
      journal       = {Phys. Rev. D},
      volume        = {84},
      pages         = {023516},
      year          = {2011},
      doi           = {10.1103/PhysRevD.84.023516},
      eprint        = {1101.2876},
      archivePrefix = {arXiv},
      primaryClass  = {hep-ph}
}

@techreport{Smith2002PCA,
    author      = {Smith, Lindsay I.},
    title       = {{A Tutorial on Principal Component Analysis}},
    institution = {Cornell University},
    address     = {Ithaca, NY, USA},
    year        = {2002},
    number      = {C. S. TR 2002-02},
    url         = {https://www.iro.umontreal.ca/~pift6080/H09/documents/papers/pca_tutorial.pdf}
}

@misc{cajohare_axionlimits,
  author       = {O'Hare, Ciaran A. J.},
  title        = {{AxionLimits}},
  howpublished = {\url{https://github.com/cajohare/AxionLimits}},
  note         = {A collection of axion and dark photon limits.}
}

@article{vectorDM,
      author         = "Graham, Peter W. and Mardon, Jeremy and Rajendran,
                        Surjeet",
      title          = "{Vector Dark Matter from Inflationary Fluctuations}",
      journal        = "Phys. Rev.",
      volume         = "D93",
      year           = "2016",
      number         = "10",
      pages          = "103520",
      doi            = "10.1103/PhysRevD.93.103520",
      eprint         = "1504.02102",
      archivePrefix  = "arXiv",
      primaryClass   = "hep-ph",
      SLACcitation   = "%%CITATION = ARXIV:1504.02102;%%"
}

@article{vectorDMKolb,
    author = "Kolb, Edward W. and Long, Andrew J.",
    title = "{Completely dark photons from gravitational particle production during the inflationary era}",
    eprint = "2009.03828",
    archivePrefix = "arXiv",
    primaryClass = "astro-ph.CO",
    doi = "10.1007/JHEP03(2021)283",
    journal = "JHEP",
    volume = "03",
    pages = "283",
    year = "2021"
}

@article{vectorDMEma,
    author = "Ema, Yohei and Nakayama, Kazunori and Tang, Yong",
    title = "{Production of purely gravitational dark matter: the case of fermion and vector boson}",
    eprint = "1903.10973",
    archivePrefix = "arXiv",
    primaryClass = "hep-ph",
    reportNumber = "DESY-19-050, DESY 19-050, KEK-TH-2114, UT-19-04",
    doi = "10.1007/JHEP07(2019)060",
    journal = "JHEP",
    volume = "07",
    pages = "060",
    year = "2019"
}

@article{vectorDMSocha,
    author = "Ahmed, Aqeel and Grzadkowski, Bohdan and Socha, Anna",
    title = "{Gravitational production of vector dark matter}",
    eprint = "2005.01766",
    archivePrefix = "arXiv",
    primaryClass = "hep-ph",
    doi = "10.1007/JHEP08(2020)059",
    journal = "JHEP",
    volume = "08",
    pages = "059",
    year = "2020"
}

@article{vectorDM1,
      author         = "Agrawal, Prateek and Kitajima, Naoya and Reece, Matthew
                        and Sekiguchi, Toyokazu and Takahashi, Fuminobu",
      title          = "{Relic Abundance of Dark Photon Dark Matter}",
      journal        = "Phys. Lett.",
      volume         = "B801",
      year           = "2020",
      pages          = "135136",
      doi            = "10.1016/j.physletb.2019.135136",
      eprint         = "1810.07188",
      archivePrefix  = "arXiv",
      primaryClass   = "hep-ph",
      reportNumber   = "TU-1074,IPMU18-015,RESCEU-13/18,MIT-CTP/5066, TU-1074,
                        IPMU18-015, RESCEU-13/18, MIT-CTP/5066",
      SLACcitation   = "%%CITATION = ARXIV:1810.07188;%%"
}

@article{vectorDM2,
    author = "Co, Raymond T. and Pierce, Aaron and Zhang, Zhengkang and Zhao, Yue",
    title = "{Dark Photon Dark Matter Produced by Axion Oscillations}",
    eprint = "1810.07196",
    archivePrefix = "arXiv",
    primaryClass = "hep-ph",
    reportNumber = "LCTP-18-21",
    doi = "10.1103/PhysRevD.99.075002",
    journal = "Phys. Rev. D",
    volume = "99",
    number = "7",
    pages = "075002",
    year = "2019"
}

@article{vectorDM3,
    author = "Dror, Jeff A. and Harigaya, Keisuke and Narayan, Vijay",
    title = "{Parametric Resonance Production of Ultralight Vector Dark Matter}",
    eprint = "1810.07195",
    archivePrefix = "arXiv",
    primaryClass = "hep-ph",
    doi = "10.1103/PhysRevD.99.035036",
    journal = "Phys. Rev. D",
    volume = "99",
    number = "3",
    pages = "035036",
    year = "2019"
}

@article{vectorDMown,
      author         = "Bastero-Gil, Mar and Santiago, Jose and Ubaldi, Lorenzo
                        and Vega-Morales, Roberto",
      title          = "{Vector dark matter production at the end of inflation}",
      journal        = "JCAP",
      volume         = "1904",
      year           = "2019",
      number         = "04",
      pages          = "015",
      doi            = "10.1088/1475-7516/2019/04/015",
      eprint         = "1810.07208",
      archivePrefix  = "arXiv",
      primaryClass   = "hep-ph",
      reportNumber   = "UG-FT 328-18, CAFPE 198-18, SISSA 44/2018/FISI",
      SLACcitation   = "%%CITATION = ARXIV:1810.07208;%%"
}

@article{Arias2012,
    author = "Arias, Paola and Cadamuro, Davide and Goodsell, Mark and Jaeckel, Joerg and Redondo, Javier and Ringwald, Andreas",
    title = "{WISPy Cold Dark Matter}",
    eprint = "1201.5902",
    archivePrefix = "arXiv",
    primaryClass = "hep-ph",
    reportNumber = "DESY-11-226, MPP-2011-140, CERN-PH-TH-2011-323, IPPP-11-80, DCPT-11-160",
    doi = "10.1088/1475-7516/2012/06/013",
    journal = "JCAP",
    volume = "06",
    pages = "013",
    year = "2012"
}

@article{Beaufort:2023qpd,
    author = "Beaufort, C. and Bastero-Gil, M. and Catalano, A. and Erfani-Harami, D-S. and Guillaudin, O. and Macias-Perez, J. and Santos, D. and Savorgnano, S. and Vezzu, F.",
    title = "{Directional detection of meV dark photons with Dandelion}",
    eprint = "2310.16505",
    archivePrefix = "arXiv",
    primaryClass = "physics.ins-det",
    doi = "10.1088/1475-7516/2024/06/058",
    journal = "JCAP",
    volume = "06",
    pages = "058",
    year = "2024"
}

@article{vectorDMRedi,
    author = "Redi, Michele and Tesi, Andrea",
    title = "{Dark photon Dark Matter without Stueckelberg mass}",
    eprint = "2204.14274",
    archivePrefix = "arXiv",
    primaryClass = "hep-ph",
    doi = "10.1007/JHEP10(2022)167",
    journal = "JHEP",
    volume = "10",
    pages = "167",
    year = "2022"
}

@article{vectorDMSato,
    author = "Sato, Takanori and Takahashi, Fuminobu and Yamada, Masaki",
    title = "{Gravitational production of dark photon dark matter with mass generated by the Higgs mechanism}",
    eprint = "2204.11896",
    archivePrefix = "arXiv",
    primaryClass = "hep-ph",
    reportNumber = "TU-1151",
    doi = "10.1088/1475-7516/2022/08/022",
    journal = "JCAP",
    volume = "08",
    number = "08",
    pages = "022",
    year = "2022"
}

@article{vectorDMCS1,
    author = "Long, Andrew J. and Wang, Lian-Tao",
    title = "{Dark Photon Dark Matter from a Network of Cosmic Strings}",
    eprint = "1901.03312",
    archivePrefix = "arXiv",
    primaryClass = "hep-ph",
    doi = "10.1103/PhysRevD.99.063529",
    journal = "Phys. Rev. D",
    volume = "99",
    number = "6",
    pages = "063529",
    year = "2019"
}

@article{vectorDMCS2,
    author = "Kitajima, Naoya and Nakayama, Kazunori",
    title = "{Dark photon dark matter from cosmic strings and gravitational wave background}",
    eprint = "2212.13573",
    archivePrefix = "arXiv",
    primaryClass = "hep-ph",
    reportNumber = "TU-1175, KEK-QUP-2022-0021",
    doi = "10.1007/JHEP08(2023)068",
    journal = "JHEP",
    volume = "08",
    pages = "068",
    year = "2023"
}

@article{vectorDMCS3,
    author = "Kitajima, Naoya and Nakayama, Kazunori",
    title = "{Nanohertz gravitational waves from cosmic strings and dark photon dark matter}",
    eprint = "2306.17390",
    archivePrefix = "arXiv",
    primaryClass = "hep-ph",
    reportNumber = "TU-1199, KEK-QUP-2023-0015",
    doi = "10.1016/j.physletb.2023.138213",
    journal = "Phys. Lett. B",
    volume = "846",
    pages = "138213",
    year = "2023"
}

@ARTICLE{Mac2024,
author = {{Mac{\'\i}as-P{\'e}rez}, J.~F. and {Fern{\'a}ndez-Torreiro}, M. and {Catalano}, A. and {Fasano}, A. and {Aguiar}, M. and {Beelen}, A. and {Benoit}, A. and {Bideaud}, A. and {Bounmy}, J. and {Bourrion}, O. and {Calvo}, M. and {Castro-Almaz{\'a}n}, J.~A. and {de Bernardis}, P. and {De Petris}, M. and {de Taoro}, A.~P. and {Garde}, G. and {G{\'e}nova-Santos}, R.~T. and {Gomez}, A. and {G{\'o}mez-Renasco}, M.~F. and {Goupy}, J. and {Hoarau}, C. and {Hoyland}, R. and {Lagache}, G. and {Marpaud}, J. and {Marton}, M. and {Masi}, S. and {Monfardini}, A. and {Peel}, M.~W. and {Pisano}, G. and {Ponthieu}, N. and {Rebolo}, R. and {Roni}, S. and {Roudier}, S. and {Rubi{\~n}o-Mart{\'\i}n}, J.~A. and {Tourres}, D. and {Tucker}, C. and {Viera-Curvelo}, T. and {Vescovi}, C.},
title = "{KISS: Instrument Description and Performance}",
journal = {Publications of the Astronomical Society of the Pacific},
year = 2024,
month = nov,
volume = {136},
number = {11},
eid = {114505},
pages = {114505},
doi = {10.1088/1538-3873/ad8189},
archivePrefix = {arXiv},
eprint = {2409.20272},
primaryClass = {astro-ph.IM},
adsurl = {https://ui.adsabs.harvard.edu/abs/2024PASP..136k4505M}
}

@ARTICLE{Monfardini2012,
author = {{Monfardini}, A. and {Benoit}, A. and {Bideaud}, A. and {Boudou}, N. and {Calvo}, M. and {Camus}, P. and {Hoffmann}, C. and {D{\'e}sert}, F.-X. and {Leclercq}, S. and {Roesch}, M. and {Schuster}, K. and {Ade}, P. and {Doyle}, S. and {Mauskopf}, P. and {Pascale}, E. and {Tucker}, C. and {Bourrion}, A. and {Macias-Perez}, J. and {Vescovi}, C. and {Barishev}, A. and {Baselmans}, J. and {Ferrari}, L. and {Yates}, S.~J.~C. and {Cruciani}, A. and {De Bernardis}, P. and {Masi}, S. and {Giordano}, C. and {Marghesin}, B. and {Leduc}, H.~G. and {Swenson}, L.},
title = "{The N{\'e}el IRAM KID Arrays (NIKA)}",
journal = {Journal of Low Temperature Physics},
year = 2012,
month = jun,
volume = {167},
number = {5-6},
pages = {834-839},
doi = {10.1007/s10909-011-0451-0},
adsurl = {https://ui.adsabs.harvard.edu/abs/2012JLTP..167..834M}
}

@ARTICLE{Bourrion2022JInst,
author = {{Bourrion}, O. and {Hoarau}, C. and {Bounmy}, J. and {Tourres}, D. and {Vescovi}, C. and {Bouly}, J.-L. and {Ponchant}, N. and {Beelen}, A. and {Calvo}, M. and {Catalano}, A. and {Goupy}, J. and {Lagache}, G. and {Mac{\'\i}as-P{\'e}rez}, J.-F. and {Marpaud}, J. and {Monfardini}, A.},
title = "{CONCERTO: readout and control electronics}",
journal = {Journal of Instrumentation},
year = 2022,
month = oct,
volume = {17},
number = {10},
eid = {P10047},
pages = {P10047},
doi = {10.1088/1748-0221/17/10/P10047},
archivePrefix = {arXiv},
eprint = {2208.07629},
primaryClass = {astro-ph.IM},
adsurl = {https://ui.adsabs.harvard.edu/abs/2022JInst..17P0047B}
}

@ARTICLE{Bourrion2016JInst,
author = {{Bourrion}, O. and {Benoit}, A. and {Bouly}, J.~L. and {Bouvier}, J. and {Bosson}, G. and {Calvo}, M. and {Catalano}, A. and {Goupy}, J. and {Li}, C. and {Mac{\'\i}as-P{\'e}rez}, J.~F. and {Monfardini}, A. and {Tourres}, D. and {Ponchant}, N. and {Vescovi}, C.},
title = "{NIKEL\_AMC: readout electronics for the NIKA2 experiment}",
journal = {Journal of Instrumentation},
year = 2016,
month = nov,
volume = {11},
number = {11},
pages = {P11001},
doi = {10.1088/1748-0221/11/11/P11001},
archivePrefix = {arXiv},
eprint = {1602.01288},
primaryClass = {astro-ph.IM},
adsurl = {https://ui.adsabs.harvard.edu/abs/2016JInst..1111101B}
}

@ARTICLE{Bounmy2022,
author = {{Bounmy}, Julien and {Hoarau}, Christophe and {Mac{\'\i}as-P{\'e}rez}, Juan-Francisco and {Beelen}, Alexandre and {Beno{\^\i}t}, Alain and {Bourrion}, Olivier and {Calvo}, Martino and {Catalano}, Andrea and {Fasano}, Alessandro and {Goupy}, Johannes and {Lagache}, Guilaine and {Marpaud}, Julien and {Monfardini}, Alessandro},
title = "{CONCERTO: Digital processing for finding and tuning LEKIDs}",
journal = {Journal of Instrumentation},
year = 2022,
month = aug,
volume = {17},
number = {8},
eid = {P08037},
pages = {P08037},
doi = {10.1088/1748-0221/17/08/P08037},
archivePrefix = {arXiv},
eprint = {2206.11554},
primaryClass = {astro-ph.IM},
adsurl = {https://ui.adsabs.harvard.edu/abs/2022JInst..17P8037B}
}

@ARTICLE{Fasano2021,
author = {{Fasano}, A. and {Mac{\'\i}as-P{\'e}rez}, J.~F. and {Benoit}, A. and {Aguiar}, M. and {Beelen}, A. and {Bideaud}, A. and {Bounmy}, J. and {Bourrion}, O. and {Bres}, G. and {Calvo}, M. and {Castro-Almaz{\'a}n}, J.~A. and {Catalano}, A. and {de Bernardis}, P. and {De Petris}, M. and {de Taoro}, A.~P. and {Fern{\'a}ndez-Torreiro}, M. and {Garde}, G. and {G{\'e}nova-Santos}, R. and {Gomez}, A. and {G{\'o}mez-Renasco}, M.~F. and {Goupy}, J. and {Hoarau}, C. and {Hoyland}, R. and {Lagache}, G. and {Marpaud}, J. and {Marton}, M. and {Monfardini}, A. and {Peel}, M.~W. and {Pisano}, G. and {Ponthieu}, N. and {Rebolo}, R. and {Roudier}, S. and {Rubi{\~n}o-Mart{\'\i}n}, J.~A. and {Tourres}, D. and {Tucker}, C. and {Vescovi}, C.},
title = "{Accurate sky signal reconstruction for ground-based spectroscopy with kinetic inductance detectors}",
journal = {Astronomy \& Astrophysics},
year = 2021,
month = dec,
volume = {656},
eid = {A116},
pages = {A116},
doi = {10.1051/0004-6361/202141419},
archivePrefix = {arXiv},
eprint = {2109.03145},
primaryClass = {astro-ph.IM},
adsurl = {https://ui.adsabs.harvard.edu/abs/2021A&A...656A.116F}
}

@ARTICLE{Calvo2013,
author = {{Calvo}, M. and {Roesch}, M. and {D{\'e}sert}, F.-X. and {Monfardini}, A. and {Benoit}, A. and {Mauskopf}, P. and {Ade}, P. and {Boudou}, N. and {Bourrion}, O. and {Camus}, P. and {Cruciani}, A. and {Doyle}, S. and {Hoffmann}, C. and {Leclercq}, S. and {Macias-Perez}, J.~F. and {Ponthieu}, N. and {Schuster}, K.~F. and {Tucker}, C. and {Vescovi}, C.},
title = "{Improved mm-wave photometry for kinetic inductance detectors}",
journal = {Astronomy \& Astrophysics},
year = 2013,
month = mar,
volume = {551},
eid = {L12},
pages = {L12},
doi = {10.1051/0004-6361/201219854},
adsurl = {https://ui.adsabs.harvard.edu/abs/2013A&A...551L..12C}
}
\end{document}